\documentstyle[prd,aps,floats]{revtex} 
\input{epsf}

\def\AJ{{\it Ap. J.} }
\def\AJL{{\it Ap. J. Lett.} }

\def\CQG{{\it Class. Quantum Gravity} }

\def\FP{{\it Fortschr. Physik} }
\def\GRG{{\it Gen. Relativity and Gravitation} }

\def\JHEP{{\it JHEP} }

\def\MPL{{\it Mod. Phys. Lett.} }
\def\MNRAS{{\it Mon. Not. R. Ast. Soc.} }
\def\NAT{{\it Nature} }
 
\def\NC{{\it Il Nuovo Cimento} }
\def\NP{{\it Nucl. Phys.} }
\def\PL{{\it Phys. Lett.} }
\def\PR{{\it Phys. Rev.} }
\def\PRL{{\it Phys. Rev. Lett.} }

\def\be{\beta}

\def\frac#1#2{{\textstyle{{#1}\over {#2}}}}

\def\lsim{\mathrel{\rlap{\lower4pt\hbox{\hskip1pt$\sim$}}
    \raise1pt\hbox{$<$}}}
\def\gsim{\mathrel{\rlap{\lower4pt\hbox{\hskip1pt$\sim$}}
    \raise1pt\hbox{$>$}}}
\def\sqr#1#2{{\vcenter{\vbox{\hrule height.#2pt
         \hbox{\vrule width.#2pt height#1pt \kern#1pt
         \vrule width.#2pt}
         \hrule height.#2pt}}}}

\def\be{\begin{equation}}
\def\ee{\end{equation}}
\def\bea{\begin{eqnarray}} 
\def\eea{\end{eqnarray}}

\begin{document}

\twocolumn[\hsize\textwidth\columnwidth\hsize\csname 
@twocolumnfalse\endcsname
\draft

\title{Generalized Chaplygin Gas, Accelerated Expansion and Dark 
Energy-Matter Unification}

\author{M. C. Bento$^{1,2}$, O. Bertolami$^1$ and A.A. Sen$^3$}

\address{$^1$  Departamento de F\'\i sica, Instituto Superior T\'ecnico \\
Av. Rovisco Pais 1, 1049-001 Lisboa, Portugal}

\address{$^2$  Centro de F\'{\i}sica das 
Interac\c c\~oes Fundamentais, Instituto Superior T\'ecnico}

\address{$^3$  Centro Multidisciplinar de Astrof\'{\i}sica, 
Instituto Superior T\'ecnico}

\address{E-mail addresses: bento@sirius.ist.utl.pt; orfeu@cosmos.ist.utl.pt; 
anjan@x9.ist.utl.pt}

\vskip 0.5cm

\date{\today}

\maketitle

\begin{abstract}
We consider the scenario emerging from the dynamics of a 
generalized $d$-brane in a $(d+1, 1)$ spacetime. The equation of state 
describing this system is given in terms of the energy density, $\rho$, and 
pressure, $p$, by the relationship $p = - A/\rho^{\alpha}$, where $A$ 
is a positive constant and $0 < \alpha \le 1$. We discuss the conditions 
under which homogeneity arises and show that this equation of state 
describes the evolution of a universe evolving from a phase dominated by 
non-relativistic matter to a phase dominated by a cosmological constant via an 
intermediate period where the effective equation of state is 
given by $p = \alpha \rho$.

\vskip 0.5cm
 
\end{abstract}

\pacs{PACS numbers: 98.80.Cq, 98.65.Es \hspace{2cm}Preprint DF/IST-2.2002, 
FISIST/04-2002/CFIF}

\vskip 2pc]
\section{Introduction}

There is mounting  evidence that the Universe at present is dominated by a 
smooth  component with negative pressure, the so-called dark energy, leading 
to accelerated expansion.
While the most obvious candidate for such component is vacuum energy, a 
plausible alternative is dynamical vacuum energy 
\cite{Bronstein,Ratra}, or quintessence. 
These models most often involve a single field 
\cite{Ratra,Wetterich,Caldwell,Ferreira,Zlatev,Binetruy,Kim,Bento,Uzan,Skordis} 
or, in some cases, two coupled fields \cite{Fujii,Masiero,Bento1}. However, 
these models usually face fine-tuning problems, notably the cosmic 
coincidence problem i.e. the question of explaining why the vacuum energy 
or scalar field dominate the Universe only recently. 
In its tracker version, quintessence models address this problem in that 
the evolution of the quintessence energy density is fairly independent of 
initial conditions; however, this seems to be achieved  at the expense of 
fine-tuning the potential parameters so that the quintessence energy density 
changes behaviour around the epoch of matter-radiation equality so as to 
overtake the matter energy density at present, driving the Universe into 
accelerated expansion.
Moreover, for quintessence models with shallow potentials, the quintessence 
field has to be nearly massless  and one expects radiative corrections to 
destabilize the ratio beteween this mass and the other known scales of 
physics; on the other hand,  the  couplings of such a light field to 
ordinary matter  give rise to  long-range forces,  which should have 
been detected in precision tests of gravity within the solar system, 
and time dependence of the constants of nature \cite{Carroll}.

Recently, it has been suggested that the  change of behavior of the missing 
energy density might be regulated by the change in the equation 
of state of the background fluid instead of  the form of the potential, 
thereby avoiding the abovementioned fine-tuning problems. This is achieved 
via the introduction, within the framework of FRW cosmology, of an exotic 
background fluid, the Chaplygin gas, described by the equation of state

\be
p = - {A \over \rho^\alpha}~~,
\label{eqstate}
\ee
with $\alpha=1$ and A a positive constant. Inserting this equation of state 
into the relativistic energy conservation equation, leads to a density evolving
as

\be
\rho =  \sqrt{A + {B \over a^6}}~~,
\label{evoldensity}
\ee 
where $a$ is the scale factor of the Universe and $B$ is an integration 
constant. This simple and elegant model smoothly
interpolates between a dust dominated phase where, 
$\rho \simeq \sqrt{B} a^{-3}$, and  
a De Sitter phase where $p \simeq - \rho$, 
through an intermediate regime described 
by the  equation of state for  stiff matter, $p = \rho$ \cite{Kamenshchik}.
Interestingly, this setup admits a well established brane interpretation as
Eq. (\ref{eqstate}), for $\alpha = 1$, is the equation of state associated 
with the parametrization invariant Nambu-Goto $d$-brane 
action in a $(d+1, 1)$ spacetime. This action leads, in the light-cone 
parametrization, to the Galileo-invariant Chaplygin gas in a $(d, 1)$
spacetime 
and to the Poincar\'e-invariant Born-Infeld action in a $(d, 1)$ spacetime 
(see \cite{Jackiw} and references therein for a thorough discussion). 
Moreover, the Chaplygin gas is the only gas known to 
admit a supersymmetric generalization \cite{Jackiw}. 

It is clear that this model has a bearing on the observed accelerated 
expansion of the Universe \cite{Perlmutter} as it automatically 
leads to an asymptotic phase 
where the equation of state is dominated by a cosmological constant, 
$8 \pi G \sqrt{A}$.    
Subsequently, it has been shown that this model admits, under conditions, 
an inhomogeneous generalization which can be regarded as a unification 
of dark matter and dark energy \cite{Bilic}.
The unification idea has received much attention recently 
\cite{Matos,Wetterich1,Kasuya,Davidson}. For instance, 
in Ref. \cite{Wetterich1}
it is suggested that  dark matter might consist of  quintessence (cosmon) 
lumps, and in Ref. \cite{Kasuya} it is shown that spintessence-like models 
are generally unstable to formation of Q-Balls which behave as 
pressureless matter.

In this work, we consider the case of a generic $\alpha$ constant in the range 
$0 < \alpha \le 1$ and show that it  
interpolates betwwen 
a universe dominated by dust and  a De Sitter one via a phase described 
by a ``soft'' matter equation of state, $p = \alpha \rho$ ($\alpha \not= 1$). 
We  show that the model can be easily accomodated in the standard 
structure formation scenarios and does not leave any undesirable signature 
on the Cosmic Microwave Background power spectrum. Furthermore, we  show 
that the 
model 
corresponds to a generalized Nambu-Goto action which can be interpreted as a
perturbed $d$-brane in a $(d+1, 1)$ spacetime.

\section{The Model}

Our starting point is the Lagrangian density for a massive complex scalar 
field, $\Phi$, 
\be
{\cal L} = g^{\mu \nu} \Phi^{*}_{, \mu} \Phi_{, \nu} - V(\vert \Phi \vert^2)~~,
\label{complexfield}
\ee
which, as suggested in Ref. \cite{Bilic},
can be expressed in terms of its masss, $m$, 
as $\Phi = ({\phi \over \sqrt{2}} m) \exp(- im \theta)$. The Lagrangian 
density (\ref{complexfield}) can then be  
rewritten as 

\be
{\cal L} = {1 \over2} g^{\mu \nu} \left(\phi^2 \theta_{, \mu} \theta_{, \nu} 
+ {1 \over m^2} \phi_{, \mu} \phi_{, \nu}\right)  - V(\phi^2/2)~~.
\label{complexfield1}
\ee 
This sets  the scale of the inhomogeneity since, assuming that 
spacetime variations of  $\phi$ correspond to scales greater than 
$m^{-1}$, then

\be
\phi_{, \mu} << m \phi~~.
\label{inhom}
\ee   
This is in contrast with  the work of Ref. \cite{Kamenshchik}, where 
spatial homogeneity 
is  assumed, and it is clearly a quite relevant contribution to generalize 
the use of the Chaplygin gas equation of state into the 
cosmological description. 
In this (Thomas-Fermi) approximation, the resulting Lagrangian density can now 
be written as 

\be
{\cal L}_{TF} = {\phi^2 \over 2} g^{\mu \nu} \theta_{, \mu} \theta_{, \nu} 
- V(\phi^2/2)~~.
\label{Thomas-Fermi}
\ee 
The corresponding equations of motion are given by

\be
g^{\mu \nu} \theta_{, \mu} \theta_{, \nu} = V'(\phi^2/2)~~,
\label{eq.motion1}
\ee

\be
(\phi^2 \sqrt{- g} g^{\mu \nu} \theta_{, \nu})_{, \mu} = 0~~,
\label{eq.motion2}
\ee
where $V'(x) \equiv dV/dx$. The field $\theta$ can be regarded as a 
velocity field provided $V'> 0$, i.e.

\be
U^{\mu} = {g^{\mu \nu} \theta_{, \nu} \over \sqrt{V'}}~~,
\label{velocity}
\ee
so that on the mass shell $U^{\mu} U_{\mu} = 1$. It then follows 
that the energy-momentum tensor built from the Lagrangian density 
Eq. (\ref{Thomas-Fermi}) takes the form of a perfect fluid whose 
thermodynamic variables can be written as 

\be
\rho = {\phi^2 \over 2} V' + V~~, 
\label{pardensity}
\ee
 
\be
p = {\phi^2 \over 2} V' - V~~. 
\label{parpressure}
\ee

Imposing the covariant 
conservation of the energy-momentum tensor for an homogeneous 
and isotropic spacetime

\be
\dot{\rho} + 3 H (p + \rho) = 0~~, 
\label{emtcons}
\ee  
where $H = \dot{a}/a$ is the expansion rate of the Universe, we get, 
for the generalized Chaplygin gas equation of state, Eq. (\ref{eqstate}), a 
generalized version of Eq. (\ref{evoldensity})

\be
\rho =  \left(A + {B \over a^{3 (1 + \alpha)}}\right)^{1 \over 1 + \alpha}~~.
\label{genevolden}
\ee 

>From Eqs. (\ref{pardensity}) and (\ref{parpressure}), we obtain

\be
d~\ln \phi^2 = {d(\rho - p) \over \rho + p}~~,
\label{phidenpres}
\ee 
which, together with Eq. (\ref{eqstate}), leads to a relationship between 
$\phi^2$ and $\rho$:

\be
\phi^2(\rho) = \rho^{\alpha} 
(\rho^{1 + \alpha} - A)^{{1 - \alpha \over 1 + \alpha}}~~.
\label{phidens}
\ee

Further algebraic manipulation, introducing Eqs. (\ref{pardensity}), 
(\ref{parpressure}) 
and (\ref{phidenpres}) into the Lagrangian density (\ref{Thomas-Fermi}),
shows that it is possible to establish a brane connection to this setting, 
as the resulting Lagrangian density has 
the form of a {\it generalized} Born-Infeld theory:

\be
{\cal L}_{GBI} = - A^{1 \over 1 + \alpha} 
\left[1 - (g^{\mu \nu} \theta_{, \mu} 
\theta_{, \nu})^{1 + \alpha \over 2\alpha}\right]^{\alpha \over 1 + \alpha}~~,
\label{GenBorn-Infeld}
\ee 
which clearly reproduces the Born-Infeld Lagrangian density for $\alpha = 1$. 
This Lagrangian density can be regarded as a $d$-brane plus soft correcting 
terms; indeed, expanding the root in Eq. (\ref{GenBorn-Infeld}) around 
$\alpha = 1$, one obtains:

\bea
\left[1 - X^{1 + \alpha \over 2 \alpha}\right]^{\alpha \over 1 + \alpha}&& = 
\sqrt{1 - X} \nonumber \\ 
&+& {X \log(X) + (1 - X) \log (1 - X) \over 4 \sqrt{1 - X}} (1 - \alpha) 
\nonumber \\
&+& {E + F + G \over 32 (1 - X)^{3/2}} (1 - \alpha)^2 + 
{\cal O}((1 - \alpha)^3)~,
\label{expansion}
\eea 
where $X \equiv g^{\mu \nu} \theta_{, \mu} \theta_{, \nu}$ and

\be
E = X (X - 2) \log^{2}(X)~~,
\label{E}
\ee

\be
F = - 2 X (X - 1) \log(X)[\log(1 - X) - 2]~~,
\label{F}
\ee
 
\be
G = (X - 1)^2 [\log(1 - X) - 4] \log(1 - X)~~.
\label{G}
\ee
 
The potential arising from this model can be written as

\be
V = {\rho^{1 + \alpha} + A \over 2 \rho^{\alpha}} = 
{1 \over 2} \left(\Psi^{2/\alpha} + {A \over \Psi^2}\right)~~,
\label{pot}
\ee
where $\Psi \equiv B^{-(1 - \alpha/1 + \alpha)} a^{3(1 - \alpha)} \phi^2$, 
which reduces to the duality invariant, $\phi^2 \rightarrow A/\phi^2$, 
and scale-factor independent potential for the 
Chaplygin gas.

The effective equation of state in the intermediate regime between the 
dust dominated 
phase and the De Sitter phase can be obtained expanding 
Eq. (\ref{genevolden}) in subleading order:

\be
\rho \simeq A^{1 \over 1 + \alpha} + \left({1 \over 1 + \alpha}\right) 
{B \over A^{\alpha \over 1 + \alpha}} a^{-3(1 + \alpha)}~~,
\label{effecden}
\ee 

\be
p \simeq - A^{1 \over 1 + \alpha} + \left({\alpha \over 1 + \alpha}\right) 
{B \over A^{\alpha \over 1 + \alpha}} a^{-3(1 + \alpha)}~~,
\label{effecpres}
\ee 
which corresponds to a mixture of  vacuum energy density  
$A^{1 \over 1 + \alpha}$ 
and matter described by the ``soft'' equation of state:

\be
p = \alpha \rho~~.
\label{effeceqstate}
\ee 

In broad terms, the comparison between the cosmological setting we propose 
and the one emerging from the Chaplygin gas, discussed in Refs. 
\cite{Kamenshchik,Bilic}, 
is exhibited in Figure 1. Naturally, a complete cosmological scenario 
involves the inclusion of radiation, 
which is related to the massless degrees of 
freedom of the Standard Model 
at a given temperature and that were dominant before recombination. 
These clearly do not affect any of the features of the scenario 
we propose here. Less trivial, 
however, is the treatment of the inhomogeneities we have allowed in 
our setting. We analyse this issue in what follows.

\begin{figure}[t]
\centering
\leavevmode \epsfysize=7cm \epsfbox{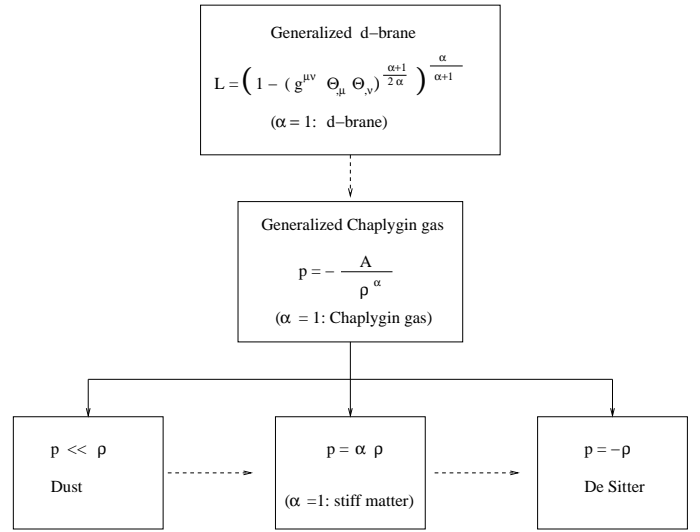}\\
\vskip 0.5cm
\caption{Cosmological evolution of a universe described by 
a generalized Chaplygin gas equation of state.}
\label{comparison2}
\end{figure}

Our starting point is
Eq. (\ref{eq.motion2}), which can be shown to 
admit as first integral a position dependent function $B(\vec{r})$, 
after a convenient choice of  comoving coordinates where the 
velocity field is given by $U^{\mu} = \delta^{\mu}_{0}/\sqrt{g_{00}}~$
\cite{Bilic}. Taking for the 
metric $g_{\mu \nu}$, the proper time $d\tau = \sqrt{g_{00}} dx^0$, and 
$\gamma \equiv -g/g_{00}$ as the determinant of the induced $3$-metric, 
then

\be
\gamma_{i j} = {g_{i0}g_{j0} \over g_{00}} - g_{ij}~~.
\label{inducedmetric}
\ee
Since for the relevant scales, function $B(\vec{r})$ can be regarded as 
approximately constant, we get

\be
\rho =  \left(A + {B \over \gamma^{(1 + \alpha)/2}}\right)^
{1 \over 1 + \alpha}~~.
\label{genevoldeninh}
\ee 

This result suggests that the Zeldovich method for considering 
inhomogeneities can be implemented through the deformation tensor 
\cite{Bilic,Matarrese,Peebles}:

\be
D_{i}^{j} = a(t) \left(\delta_{i}^{j} - b(t) 
{\partial^2 \varphi(\vec{q}) \over \partial q^i \partial q^j}\right)~~,
\label{defromationt}
\ee
where $\vec{q}$ are generalized Lagrangian coordinates so that 

\be
\gamma_{i j} = \delta_{mn}D_{i}^{m}D_{j}^{n}~~,
\label{inducedmetric1}
\ee     
and $h$ is a perturbation

\be
h = 2 b(t) {\varphi_{,i}}^i~~,
\label{hevol}
\ee
with $b(t)$ parametrizing the time evolution of the inhomogeneities.
Hence, using Eqs. above and Eqs. (\ref{effecden}) - 
(\ref{effecpres}),  it follows that

\be
\rho \simeq \bar \rho (1 + \delta)~~~,~~~p \simeq - 
{A \over \bar \rho^{\alpha}} (1 - \alpha \delta)~~,
\label{pert}
\ee 
where $\bar \rho$ is given by Eq. (\ref{genevolden}) and the density contrast, 
$\delta$, is related to $h$ through 

\be
\delta =  {h \over 2} (1 + w)~~,
\label{delta}
\ee 
and $w$ reads

\be
w \equiv {p \over \rho} = - {A \over \bar \rho^{1+ \alpha}}~~.
\label{omega}
\ee 

The metric (\ref{inducedmetric1}) leads 
to the following $0-0$ component of the Einstein equations:

\be
- 3 {\ddot{a} \over a} + {1 \over 2} \ddot{h} + H \dot{h} = 
4 \pi G \bar \rho [(1 + 3 w) 
+ (1 - 3 \alpha w) \delta]~~,
\label{Einstein}
\ee 
where the unperturbed part of this equation corresponds 
to the Raychaudhuri equation

\be
- 3 {\ddot{a} \over a} = 4 \pi G \bar \rho (1 + 3 w)~~.
\label{Raychaudhuri}
\ee 

Using the Friedmann equation for a flat spacetime

\be
H^2= {8 \pi G \over 3} \bar \rho~~,
\label{Friedmann}
\ee
Eq. (\ref{Einstein}) can be written as a differential equation for $b(a)$:

\be
{2 \over 3} a^2 b'' + (1 - w) a b' - (1 + w) (1 - 3 \alpha w) b = 0~~,
\label{bprimes}
\ee
where the primes denote derivatives with respect to the scale-factor, $a$.

\begin{figure}[t]
\centering
\leavevmode \epsfysize=7cm \epsfbox{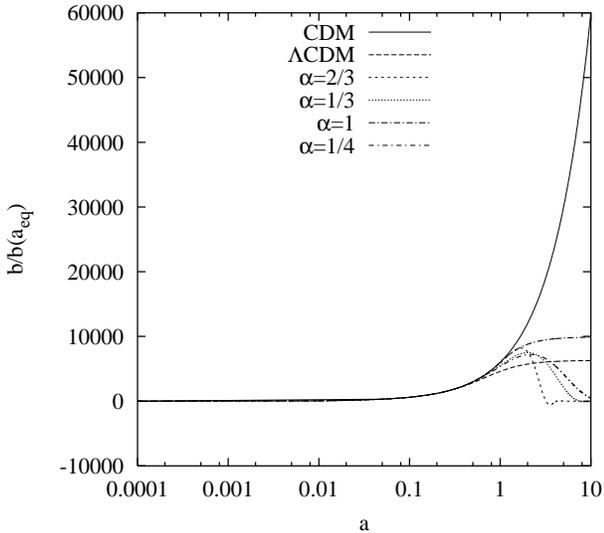}\\
\vskip 0.5cm
\caption{Evolution of $b(a)/b(a_{eq})$ for the generalized Chaplygin 
gas model, for different values 
of $\alpha$, as compared with CDM and $\Lambda$CDM.}
\label{fig:b1}
\end{figure}

\begin{figure}[t]
\centering
\leavevmode \epsfysize=7cm \epsfbox{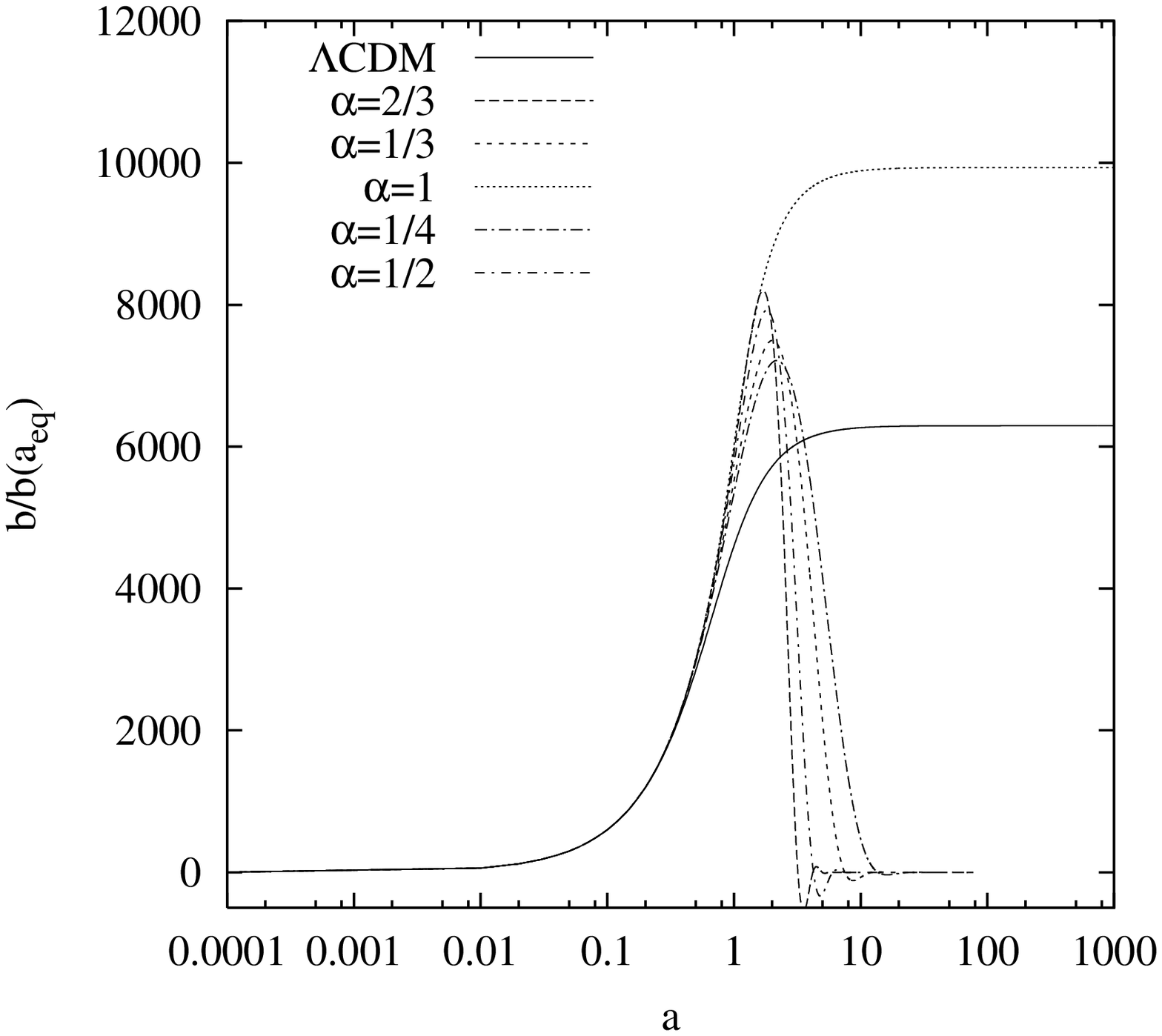}\\
\vskip 0.5cm
\caption{Evolution of $b(a)/b(a_{eq})$ for the generalized Chaplygin 
gas model, for different values of 
$\alpha$, as compared with $\Lambda$CDM.}
\label{fig:b2}
\end{figure}

Finally, from Eqs. (\ref{genevolden}) and (\ref{omega}), we derive 
an expression for $w$ as a function of the scale-factor, using 
the observational 
input that $\Omega_{\Lambda} + \Omega_{M} = 1$, 
where $\Omega_{\Lambda} \simeq 2/3$ and $\Omega_{M} \simeq 1/3$ 
\cite{Bahcall} are, respectively, the fractional vacuum
and matter (dark + baryons) energy densities

\be
w(a) = - {\Omega_{\Lambda} a^{3(1 + \alpha)} \over 1 - \Omega_{\Lambda} + 
\Omega_{\Lambda} a^{3(1 + \alpha)}}~~.
\label{fomega}
\ee   

We have used this expression to  integrate Eq. (\ref{bprimes}) numerically,
for different values of $\alpha$.  We have set $a_{eq} = 10^{-4}$ 
for matter-radiation equilibrium and $a_{0} = 1$  at present, taking 
as initial condition $b'(a_{eq}) = 0$. Our results are shown in Figures 
\ref{fig:b1} and \ref{fig:b2}.

We find that 
generalized Chaplygin scenarios start differing from the $\Lambda$CDM  
only recently ($z \simeq 1$) and that, in any case, they yield a 
density contrast that closely 
resembles, for any value of $\alpha \not= 0$, the standard CDM 
before the present. 
Notice that the $\Lambda$CDM corresponds effectively 
to setting  $\alpha = 0$ in Eq. (\ref{fomega}) 
and removing the  
factor $1 - 3 \alpha w$ in Eq. (\ref{bprimes}). Figure \ref{fig:b2}  
shows also that, for any value of $\alpha$, $b(a)$ saturates 
as in the  $\Lambda$CDM case.

\begin{figure}[t]
\centering
\leavevmode \epsfysize=7cm \epsfbox{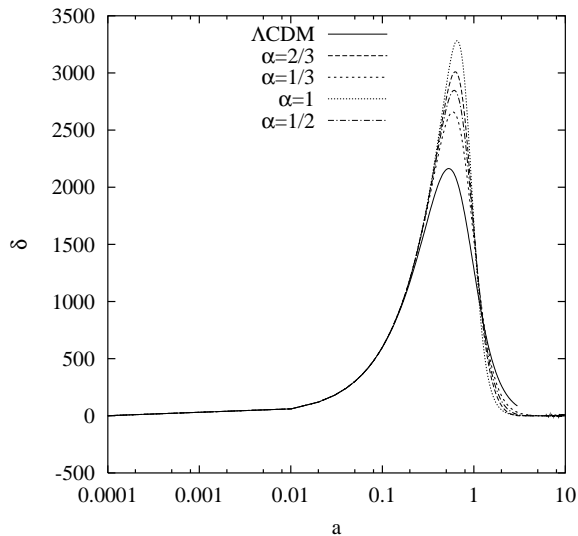}\\
\vskip 0.5cm
\caption{Density contrast for different values of $\alpha$, 
as compared with $\Lambda$CDM.}
\label{fig:pert}
\end{figure}

As for the density contrast, $\delta$, 
we can see, using Eqs. (\ref{hevol}), (\ref{delta}) and (\ref{fomega}), 
that the ratio between this quantity in the Chaplygin and the $\Lambda$CDM 
scenarios is given by:

\be
{\delta_{Chap} \over \delta_{\Lambda CDM}} = {b_{Chap} \over 
b_{\Lambda CDM}} {1 - \Omega_{\Lambda} + \Omega_{\Lambda} a^{3} \over 
1 - \Omega_{\Lambda} + \Omega_{\Lambda} a^{3(1 + \alpha)}}~~,
\label{ratio}
\ee 
meaning that their difference diminishes as $a$ evolves. In Figure 
\ref{fig:pert}, we have plotted $\delta$  as a function of $a$ for different 
values of $\alpha$; hence, we verify for any $\alpha$ the claim of 
Refs. \cite{Bilic,Fabris}, for $\alpha = 1$, that 
the density contrast decays for large $a$.
Figure \ref{fig:pert} also shows the main difference in behaviour of the 
density contrast between a 
Universe filled with matter with a ``soft'' or ``stiff'' equations of state 
as the former resembles more closely the $\Lambda$CDM.

\section{Discussion and Conclusions}

In this work, we have considered a generalization of the 
Chaplygin equation of state, $p = - {A \over \rho^\alpha}$, with 
$0 < \alpha \le 1$. We have shown that, as in the 
case of the Chaplygin gas, where $\alpha = 1$, the model admits a $d$-brane 
connection
as its Lagrangian density corresponds to 
the Born-Infeld action plus some soft logarithmic corrections. 
Furthermore, spacetime is shown to evolve 
from a phase that is initially dominated, 
in the absence of other degrees of freedom on the brane, 
by non-relativistic matter to a phase that is 
asymptotically De Sitter. This behaviour is similar to one of 
the Chaplygin gas. The intermediate 
regime in our model corresponds to a phase where the effective equation 
of state is 
given by $p = \alpha \rho$. We have estimated the fate of the inhomogeneities 
admitted in the model and shown that these evolve consistently 
with the observations as the density contrast they introduce is 
smaller than the one typical of CDM scenarios and closer to the ones predicted 
by the $\Lambda$CDM in comparison to the Chaplygin $\alpha = 1$ case.

Hence, given the 
fundamental nature of the underlying physics behind the Chaplygin gas 
and its generalizations, it appears that it contains 
some of the key ingredients in the description of the Universe 
dynamics at early as well as late times.

\begin{acknowledgments}

\noindent
M.C.B. and  O.B. acknowledge the partial support of Funda\c c\~ao para a 
Ci\^encia e a Tecnologia (Portugal)
under the grant POCTI/1999/FIS/36285. The work of A.A. Sen is fully 
financed by the same grant. We also would like to thank the authors of Ref. 
\cite{Bilic} for their relevant comments and suggestions.

\end{acknowledgments}

\end{document}